# Coexistence of unconventional spin-orbit torque and in-plane Hall effect in a single ferromagnetic layer

Jiaxin Chen[1], Hongsheng Zheng[2], Hongliang Chen[1], Qia Shen[1], Chang Pan[4], Zhenyi Zheng[3], Hemian Yi[1], Dandan Guan[1], Xiaoxue Liu[1], Yaoyi Li[1], Shiyong Wang[1], Hao Zheng[1], Canhua Liu[1], Jinfeng Jia[1], Jingsheng Chen[3], Ruidan Zhong[1], Lei Wang[5], Xuepeng Qiu[4], Yumeng Yang[2*], Aurélien Manchon[6*], Liang Liu[1,7*]

[1]*Tsung-Dao Lee Institute, Key Laboratory of Artificial Structures and Quantum Control (Ministry of Education), School of Physics and Astronomy, Shanghai Jiao Tong University, Shanghai 200240, China*
[2]*Shanghai Engineering Research Center of Energy Efficient and Custom AI IC, School of Information Science and Technology, Shanghai Tech University, Shanghai 201210, China*
[3]*Department of Materials Science and Engineering, National University of Singapore, 117575, Singapore*
[4]*Shanghai Key Laboratory of Special Artificial Microstructure Materials and Technology and School of Physics Science and Engineering, Tongji University, Shanghai 200092, China*
[5]*Key Laboratory of Quantum Materials and Devices of Ministry of Education, School of Physics, Southeast University, Nanjing 211189, China*
[6]*Aix-Marseille Université, CNRS, CINaM, Marseille, France*
[7]*Hefei National Laboratory, Hefei, China*

*e-mail: yangym1@shanghaitech.edu.cn, liul21@sjtu.edu.cn

**The symmetry of a material fundamentally governs its spin transport properties. While unconventional spin transport phenomena have been predominantly explored in low-symmetry systems (e.g., $C_{1v}$ symmetry), high-symmetry crystals—which constitute the majority of industry-compatible materials—are generally expected to exhibit only conventional spin-transport behavior. Here, we report the coexistence of two unconventional spin transport effects, the crystal spin-orbit torque (CSOT) and the crystal in-plane Hall effect (CIHE), in a CoPt single ferromagnetic layer with $C_{3v}$ symmetry. Leveraging the CSOT, we achieve nearly 100% field-free perpendicular magnetization switching in a 6 nm CoPt layer at room temperature. Simultaneously, the**

**CIHE observed in this material exhibits nearly identical dependencies on both current angle and growth temperature as the CSOT. Symmetry analysis confirms that both effects share a common physical origin. Our work not only establishes CoPt as a high-performance spin-orbit material, but also demonstrates that unconventional spin transport can be realized in high-symmetry systems, thereby opening a broad pathway for their application in practical spintronics.**

The relationship between crystal symmetry and spin transport governs the fundamental physics of spintronic phenomena for crystalline thin films and van der Waals materials[1,2]. Spin-related observables, e.g. spin conductivity, must remain invariant under the symmetry operations of a crystal[3,4]. In systems with higher symmetries, the spin transport quantities are restricted, with only limited components expected in the conductivity tensor, which represents a "conventional" behavior in spin transport. For example, only in-plane SOT is allowed in a polycrystalline heavy metal/ferromagnet (HM/FM) bilayer with $C_{nv}$ symmetry[5,6] ($n$-fold rotation along $z$-axis and $n$ vertical mirror planes, $n\to\infty$), where the mirror symmetry in the current-magnetization ($I$-$M$) plane restricts the spin polarization ($\sigma$) to the in-plane direction and orthogonal to the current flow direction[2,7]. This conventional SOT cannot deterministically switch the perpendicular magnetization, which is a major obstacle for its applications. In contrast, a reduced symmetry generally enables unconventional spin transport phenomena. For instance, in systems with only a mirror symmetry ($C_{1v}$ symmetry), previous studies have revealed out-of-plane spin polarization[8,9], nonlinear Hall effect[10,11], nonlinear magnetoresistance[12], and field-free magnetization switching[9,13-15]. However, materials with $C_{1v}$ symmetry only represent a small portion of the family consisting of 32 non-magnetic point groups[16], while the industry-compatible spintronics materials (e.g., 3d metals and alloys) usually possess higher symmetries (e.g., $C_{3v}$, $D_{4h}$). Previously, the crystal-symmetry dependent SOT[17-21] (CSOT) has been proposed in high-symmetry materials with $C_{3v}$ point group. This CSOT is also called "anisotropy-like SOT"[1] because its strength is strongly dependent on the current-flowing direction with respect to the crystal plane, in contrast to conventional damping-like SOT which is isotropic.

On the other hand, the exploration of crystal symmetry-driven transport phenomena has unveiled another unconventional effect, the crystal in-plane Hall effect (CIHE), where an in-plane magnetic field generates a transverse voltage in the materials[22-26]. Distinct from the conventional planar Hall effect (PHE), which is a manifestation of anisotropic magnetoresistance and is an even function of the magnetic field ($H$), the CIHE is a genuine Hall effect, with its signal being odd with respect to the in-plane $H$. In recent years, the CIHE has been observed in several materials, and its physical origins are multifaceted, including the Lorentz force acting on anisotropic Fermi surfaces in rutile oxides like RuO$_2$[22], the momentum-space Berry curvature in topological materials such as VS$_2$-VS superlattices[23] and Co$_3$Sn$_2$S$_2$[25], and the octupole in the magnetization space of common ferromagnets like Fe and Ni[27].

Despite the very similar crystal symmetry constraints of CSOT and CIHE, a direct observation of their coexistence remains elusive. More importantly, the microscopic origin of these two effects and their internal link remain to be clarified, especially given the diversity of interpretations that have been proposed for them. In addition, the key performance characteristics of the two effects are important for practical applications, including their strengths, working temperature, and growth conditions, which demands a systematic investigation. In this work, we report the coexistence of the CSOT and CIHE in epitaxial CoPt single layer with C$_{3v}$ symmetry. Both effects share the same crystal-symmetry dependence and growth temperature ($T_{grow}$) dependence, indicative of their similar origins. At an optimized $T_{grow}$, the CSOT enables a 100% field-free perpendicular magnetization switching, while the largest field-free switching ratio in previous single-layer SOT systems is below 60%. Concurrently, the CIHE is observed, exhibiting a high coefficient of 1.57×10$^{-3}$ cm$^3$/C. Our theoretical analysis demonstrates that the two unconventional spin transport effects originate from the trigonal warping at the Fermi surface. These findings may stimulate future investigation of unconventional spin transport in high-symmetry crystals.

**Results**

**Unconventional crystal spin-orbit torque and in-plane Hall effect in C$_{3v}$ ferromagnets**

Figure 1 depicts a magnetic system with C$_{3v}$ crystal symmetry, including threefold rotational symmetry along the *z*-axis and three vertical mirror planes. In the presence of out-of-plane magnetization ($M_z$), the C$_{3v}$ crystal possesses a warped Fermi surface (called "trigonal warping") as shown in Fig. 1b. This can induce unconventional spin transports, including CSOT[20] (or "3*m* torque"[18]) and CIHE[28] (or "chiral Hall effect"[28]), as illustrated in Fig. 1c and 1d, respectively. Based on a symmetry analysis of the system (see Supplementary Section 2), the expression of the current-induced SOT effective field and in-plane Hall resistivity can be derived. We identified a C$_{3v}$-specific term in the spin-orbit torque effective field:

$$\boldsymbol{H}_{3m} = \chi_{3m}\boldsymbol{m} \times [(m_y E_x + m_x E_y)\boldsymbol{x} + (m_x E_x - m_y E_y)\boldsymbol{y}], \quad (1)$$

named as CSOT (or "3*m* torque"[18]), and a corresponding term in the Hall current:

$$j_H = \sigma_\chi \cos(\varphi_M + 2\varphi_E) \sin\theta \cos\theta, \quad (2)$$

named as CIHE (or "chiral Hall effect"[28]).

We found that both effects are strongly dependent on the electrical current direction ($\varphi_E$). For the SOT, under the condition that the damping-like torque drives the magnetization towards the spin polarization direction (see Supplementary Section 2 for the details), the CSOT effective field simplifies to:

$$\boldsymbol{H}_{3m} = H_0 \cos(3\varphi_E)\,\boldsymbol{z}, \qquad (3)$$

where an apparent 3-fold angular dependence on $\varphi_E$ is obtained. Similarly, when the electrical current is parallel to the magnetic field ($\varphi_E = \varphi_M$), the simplified form of CIHE resistivity is given by (see Supplementary Section 2 for the details):

$$\rho_{xy}^\chi = \rho_{xy}^{\chi,0} \cos 3\varphi_E. \qquad (4)$$

It is clear that the above two effects share the same threefold angular dependence. However, their coexistence in the same system has not been experimentally observed prior to this work.

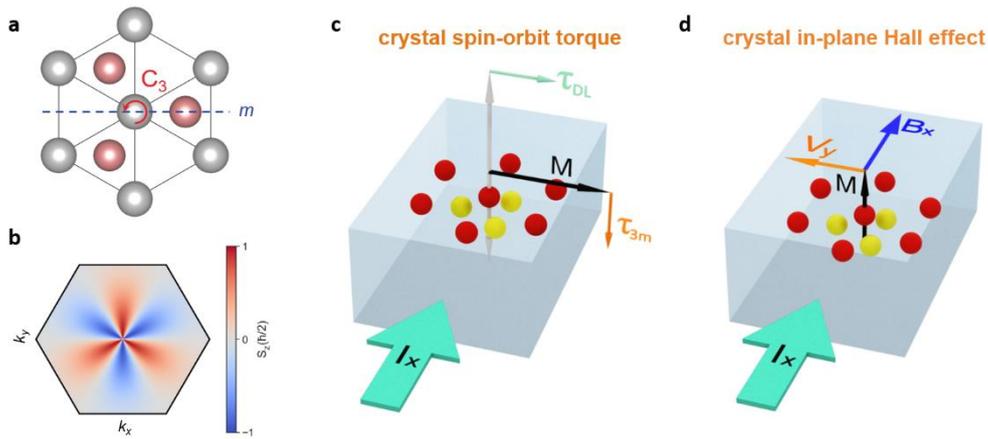

**Figure 1| crystal spin orbit torque (CSOT) and crystal in-plane Hall effect (CIHE) in C$_{3v}$ magnets. a.** Schematic of C$_{3v}$ crystal symmetry. **b.** Schematic of trigonal warping in the $k$-space of a C$_{3v}$ magnets. **c.** Schematic of crystal spin orbit torque when current is applied in the low-symmetry direction. **d.** Schematic of the crystal in-plane Hall effect when current flows in the low-symmetry direction.

**Sample preparation and characterization of CoPt films with C$_{3v}$ symmetry**

In our experiments, CoPt films with a thickness of 6 nm were epitaxially grown on MgO (111) single-crystal substrates by DC magnetron co-sputtering of Co and Pt targets at deposition temperature ($T_{\text{grow}}$) from 225 °C to 675 °C (see Methods for the details). In a composition ratio of Co$_{33}$Pt$_{67}$ (denoted as "CoPt" hereafter for simplicity), a face-centered-cubic (*fcc*) A1 phase with perpendicular magnetic anisotropy can be obtained due to the formation of Co platelet with (111) texture near the substrate. Figs. 2a-c show typical structural and magnetic characterizations of a CoPt film deposited at 350 °C. The high-resolution X-ray diffraction (HR-XRD) $\theta$-$2\theta$ scan (Fig. 2a) confirmed the (111) texture of the CoPt film, without other obvious textures. The XRD $\varphi$-scan pattern with CoPt (002) plane rotated along [111] axis exhibits a threefold rotational symmetry, revealing a twin-free epitaxial growth of (111)-oriented CoPt (see Supplementary Fig. S1). To further check the crystal structure, we performed High-angle annular dark-field scanning transmission electron microscopy (HAADF-STEM) imaging, which shows an ABCABC… stacking, consistent with a *fcc* structure[29]. The high-temperature grown CoPt films usually possess a composition gradient along the film thickness direction due to the different diffusion rates of Co and Pt[30-33]. For the sample deposited at 350 °C, the energy-dispersive X-ray spectroscopy (EDS) confirmed its composition gradient ($G$) along the growth axis with a value of $G$ = 2.4% Pt per nm. This composition gradient lowers the crystal symmetry from the -3$m$ point group with inversion symmetry to the 3$m$1 point group. XRD results for CoPt films grown at temperature ranging from 225 °C to 675 °C are shown in Supplementary Fig. S2, where similar patterns are revealed to the 350 °C sample. The basic magnetic properties of the CoPt films are shown in Supplementary Fig. S3. All of their magnetic hysteresis (*M-H*) loops along the out-of-plane direction are square, indicating perpendicular magnetic anisotropy. The saturation magnetization ($M_s$) remains stable (in the range between 460 and 500 emu/cc) with $T_{grow}$ (Fig. 2d), which indicates that the composition ratio of Co/Pt is nearly independent with $T_{grow}$. In contrast, the anisotropic field ($H_k$) shows a small fluctuation with $T_{grow}$ (Fig. 2d). We also measured the resistivity-temperature ($\rho$-$T$) curves (see Supplementary Fig. S4) and found that the room temperature $\rho$ remains almost unchanged with $T_{grow}$.

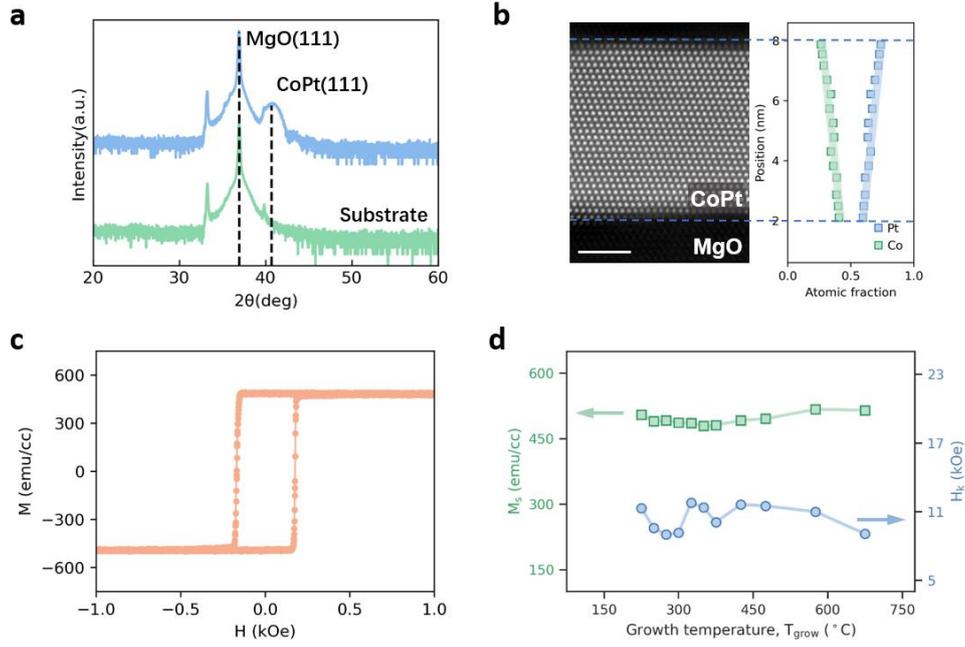

**Figure 2| Structural and magnetic characterizations of CoPt films. a.** $\theta$-$2\theta$ High-Resolution X-ray diffraction (HR-XRD) pattern of 6 nm CoPt films deposited at 350 °C. **b.** High-angle annular dark-field scanning transmission electron microscopy (HAADF-STEM) image of the CoPt (6 nm) film deposited at 350 °C. The scale bar in the figure is 2 nm. Energy-dispersive X-ray spectroscopy (EDS) elemental mapping of Co and Pt shows a decrease (increase) in Co (Pt) content along the +z direction. **c.** Out-of-plane *M-H* loop of CoPt thin film deposited at 350 °C. **d.** Growth temperature dependence of the saturation magnetization ($M_s$) and magnetic anisotropy field ($H_k$).

**Crystal spin-orbit torque and field-free switching**

To study the symmetry-dependent spin transport, we patterned the CoPt films into Hall bar devices with different current angles ($\varphi_I$) relative to the [1$\bar{1}$0] crystal axis (inset in Fig. 3a). The anomalous Hall effect (AHE) versus out-of-plane magnetic field ($R_{xy}$-$H_z$) for different $\varphi_I$ is shown in Supplementary Fig. S5, all of which show square loops. To characterize the SOT, we applied a pulsed electrical current with a 30 µs duration to drive the magnetization switching (see Methods). Figure 3a shows the current-induced field-free magnetization switching loop ($R_{xy}$-$I_{\text{pulse}}$) for a device with $\varphi_I = 0°$ and $T_{\text{grow}} = 350$ °C. The switching is achieved without the assistance of an external magnetic field, as reported in our previous work[29]. We note that in our previous work, the switching ratio ($r_{\text{sw}} = \Delta R_{xy}^I / \Delta R_{xy}^H$) of $Co_{30}Pt_{70}$ was about 54%, where $\Delta R_{xy}^H$ and $\Delta R_{xy}^I$ represent the AHE amplitude and the current-switched Hall resistance amplitude, respectively. We found that $r_{\text{sw}}$ shows a strong dependence on $T_{\text{grow}}$. Specifically, as $T_{grow}$ increases from 225 °C to 675 °C, $r_{sw}$ first shows an increasing trend and then decreases, reaching a maximum of 98% at the $T_{grow}$ of 350 °C (Fig. 3b and Supplementary Fig. S9). To investigate why the switching ratio is optimized at 350 °C, we characterized the current-induced in-plane and out-of-plane SOT effective fields. As described in Supplementary Section 4 and previous works[34, 35], we measured second harmonic Hall voltage under an AC current ($I_{ac}$) to estimate the strength of the conventional damping-like SOT effective field ($H_{DL}$). As shown in Supplementary Fig. S12, $H_{DL}$ shows little variation with $T_{grow}$, with a normalized average $H_{DL}/J$ of ~17.54 Oe/($10^7$ A/cm$^2$) ($J$ is the current density of $I_{ac}$). For comparison, we also characterized the out-of-plane effect field ($H_{OOP}$) through AHE loop shift measurements[36, 37]. The inset of Fig. 3c shows the AHE loops under pulsed currents of +35 mA and -35 mA for the 350 °C deposited sample ($\varphi_I=0°$), from which we can extract a $H_{OOP}$ of 50 Oe (see Supplementary Section 4 for the details). In contrast to the nearly constant $H_{DL}/J$, Figure 3c shows that $H_{OOP}/J$ first increases and then decreases with $T_{grow}$. At around 350 °C, $H_{OOP}/J$ is maximized to 6.99 Oe/($10^7$ A/cm$^2$). The similar $T_{grow}$ dependence of the switching ratio and the OOP effective field suggests that the CSOT is optimized at 350 °C. For further verification, we fabricated Hall bars with different $\varphi_I$ from the films grown at 350 °C. We found that $r_{sw}$ exhibits a clear threefold angular dependence (Fig. 3d), maximized when the current flows along low-symmetry axes (e.g., $\varphi_I$ = 0°, 60°, 120°, 180°) and nearly completely suppressed along high-symmetry mirror axes (e.g., $\varphi_I$ = 30°, 90°, 150°). Importantly, the $r_{sw}$-$\varphi_I$ relation can be fitted by a $\cos(3\varphi_I)$ formula.

To characterize the angular dependence of the CSOT strength, we measured the $H_{OOP}$ at various $\varphi_I$, and the details can be found in Supplementary Section 5. Then we derive the $\varphi_I$ dependence of $H_{OOP}/J$ (see Fig. 3e), which also exhibits a threefold angular dependence. The $H_{OOP}/J$-$\varphi_I$ relation can be well fitted by Eq. (3), from which we obtained the strength of the CSOT effective field: $H_0$ = 6.99 Oe/($10^7$ A/cm$^2$).

In Fig. 3f, we list two key parameters, switching ratio ($r_{sw}$) and working temperature ($T_w$), for existing single-layer systems that exhibit field-free SOT switching phenomenon. Up to now, a great effort has been made to realize the field-free SOT switching in single ferromagnetic layer by using the following strategies: tilted anisotropy[38, 39], in-plane magnetization[40], gradient DMI[41], Octahedral rotation[42, 43], and low crystal symmetry[29, 44, 45]. None of these works can realize a full field-free SOT switching at room temperature (300 K). In our work, the field-free switching ratio reaches nearly 100% at 300 K.

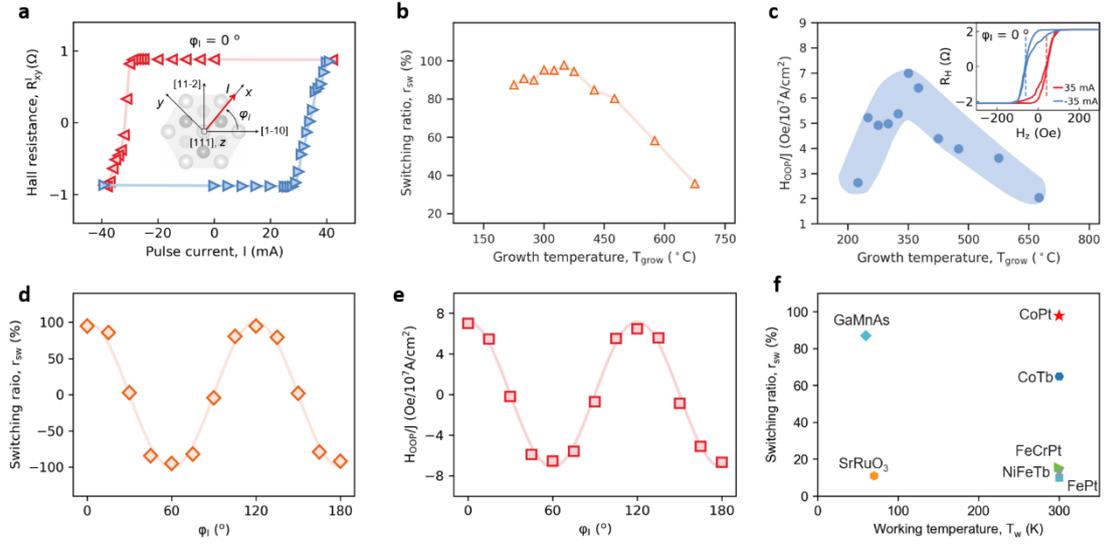

**Figure 3| Crystal spin orbit torque (CSOT) induced full field-free magnetization switching in CoPt single layer. a.** Field-free deterministic switching of perpendicular magnetization when a current pulse is applied along the $\varphi_I = 0°$ direction. **b.** Growth temperature ($T_{grow}$) dependence of the switching ratio ($r_{sw}$). **c.** Growth temperature dependence of the out-of-plane (OOP) effective field. The inset shows a representative loop shift measurement. **d, e.** Threefold angular dependence of the switching ratio (**d**) and OOP effective field (**e**) on the current direction $\varphi_I$ for the 350 °C deposited sample. Solid curves represent fits to a $\sim \cos(3\varphi_I)$ function. **f.** Comparison of the perpendicular magnetization switching ratio and switching temperature of our CoPt film with other single-layer systems.

**Crystal in-plane Hall effect**

As we discussed, the $C_{3v}$ magnets may also allow for the generation of an CIHE, which originates from the chiral Hall term in the Hall conductivity tensor[28]. To detect this effect, we measured the Hall response under a small in-plane magnetic field ($H_x$) applied along the current direction. The measurement was performed by first saturating the sample magnetization ($M$) along an out-of-plane direction (+$z$ or -$z$), to a +$M$ or -$M$ state. Subsequently, we sweep $H_x$ along the current direction and measure the total Hall resistance ($R_{xy}$). Figures 4a, 4b and Supplementary Fig. S15 present the $R_{xy}$-$H_x$ relation for both +$M$ and -$M$ states in a sample grown at 350 °C with different $\varphi_I$ from 0° to 165°. We found that for high-symmetry axes (e.g., $\varphi_I = 30°$ and 90°), the $R_{xy}$-$H_x$ relationship exhibits a typical parabolic dependence, which arises from the anomalous Hall effect ($R_{xy}^{AHE}$), indicating a coherent rotating process of the magnetization around its initialization state. In contrast, for the low-symmetry axes (e.g., $\varphi_I = $ 0° and 60°), the $R_{xy}$-$H_x$ contains both $R_{xy}^{AHE}$ and an additional term that is linearly proportional to the in-plane field $H_x$, which we identify as the CIHE resistance ($R_{xy}^\chi$). The total Hall resistance is therefore described by:

$$R_{xy} = R_{xy}^{AHE} + R_{xy}^\chi \qquad (5)$$

By fitting the Hall resistance using Eq. (5) (see more details in Supplementary Section 6), we can derive the magnitude of the CIHE coefficient, $\rho_{xy}^\chi/H_x$, where $\rho_{xy}^\chi$ is the resistivity obtained by multiplying $R_{xy}^\chi$ by the film thickness ($t$). This coefficient also exhibits a clear threefold angular dependence on $\varphi_I$ (Fig. 4c), similar to that of the CSOT. Furthermore, the $\rho_{xy}^\chi/H_x$-$\varphi_I$ relation can be well fitted by Eq. (4) as shown in Fig. 4c. We also investigated the $T_{grow}$ dependence of $R_{xy}^\chi/H_x$ as shown in Fig. 4d. We found that it reaches a maximum at 350 °C, which shows a trend similar to that of the CSOT effective field in Fig. 3c.

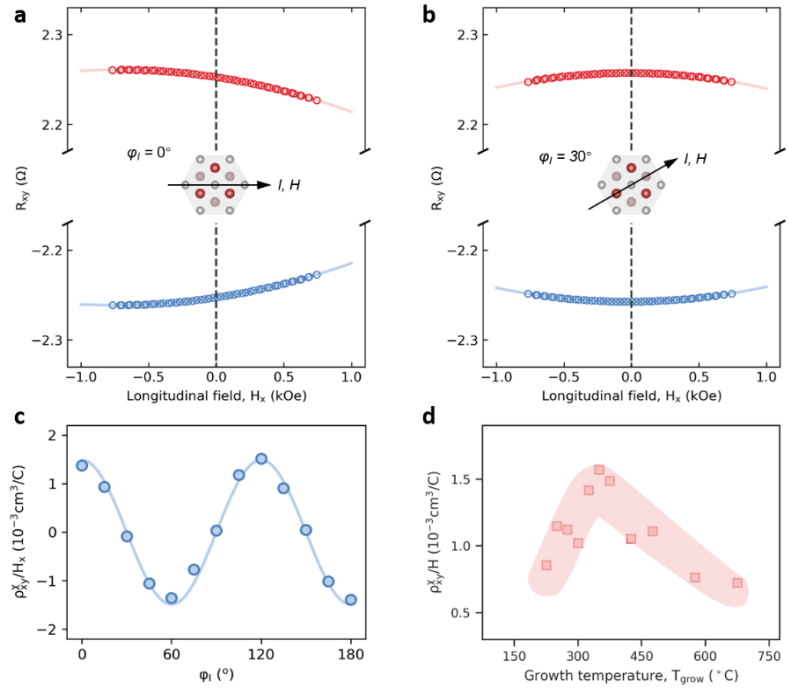

**Figure 4| Crystal in-plane Hall effect (CIHE) in CoPt single layer. a.** CIHE when current and magnetic field $H_x$ are applied along the $\varphi_I = 0°$ direction. Inset: schematic diagram of CIHE measurement. Both the current and the magnetic field are in-plane and have an azimuth angle $\varphi_I = \varphi_H$ with respect to the [1−10] direction. **b.** CIHE when current and magnetic field $H_x$ are applied along the $\varphi_I = 30°$ direction. **c.** Current angle ($\varphi_I$) dependence of the CIHE coefficient $\rho_{xy}^\chi/H_x$. The solid line represents a fitting by using Eq. (4). **d.** Growth temperature dependence ($T_{grow}$) of $\rho_{xy}^\chi/H_x$ of CoPt films.

**Discussion**

The angular dependence of both the CSOT and CIHE can be well described by the symmetry analysis and theoretical calculation in our previous work[18, 28]. This study demonstrates that a single ferromagnetic CoPt layer with $C_{3v}$ symmetry can host multiple unconventional spin transport effects, notably the CSOT and the CIHE. A central finding is the realization of nearly complete field-free perpendicular magnetization switching in a 6 nm-thick CoPt film, which, among reported single-layer ferromagnetic systems, represents a switching ratio at the upper limit of what has been achieved to date. Evaluations of thermal stability (Supplementary Section 8) indicate that the CoPt system exhibits favorable properties compared to conventional CoFeB/MgO systems. Although its energy efficiency does not surpass that of certain material systems, the high compositional and structural compatibility of CoPt with existing CMOS fabrication processes enhances its potential for practical integration. In addition, the sizable CIHE signal observed here offers a useful platform for further investigating the relationship between crystal symmetry and transverse charge-spin conversion mechanisms.

The simultaneous observation of CSOT and CIHE in CoPt highlights the critical, and often overlooked, role of point group symmetry in enabling unconventional spin transport even in high-symmetry crystals. Both effects exhibit the same threefold angular dependence and similar scaling with growth temperature, pointing to a common origin in the symmetry-broken electronic structure, specifically the trigonal warping of the Fermi surface. Beyond the effects studied here, symmetry analysis suggests that other responses, such as the nonlinear Hall effect, may also be accessible in $C_{3v}$ systems under appropriate conditions. These results illustrate that high-symmetry materials are not restricted to conventional spin behaviors; rather, their intrinsic crystal symmetry can be harnessed to realize distinct and functionally useful transport phenomena. This perspective broadens the scope for designing spintronic functionalities in material systems that are both structurally robust and technologically compatible.

**Conclusion**

In summary, we have demonstrated the coexistence of two unconventional spin transport effects in CoPt single layers: the CSOT and the CIHE. Their identical threefold angular dependence and growth temperature dependence should be due to the similar symmetry constraint of the response tensor of these two effects. The key lies in the $C_{3v}$ point group symmetry of the epitaxially grown CoPt (111) film, supported by symmetry analysis based on invariant theory. Remarkably, a 100% field-free switching was achieved at 6 nm $Co_{33}Pt_{67}$ with an optimized deposition temperature of 350 °C, which overcome a longstanding challenge of incomplete switching in single-layer field-free SOT switching systems. In principle, unconventional spin transport with specific non-zero component in the response tensor should also be allowed in other materials with specific point group symmetries, which deserves further investigations.

**Methods**

*Sample growth and device fabrication.* CoPt single layers were epitaxially deposited on single-crystal MgO (111) substrates using DC magnetron sputtering (AJA) in a system with a base pressure below $4 \times 10^{-8}$ Torr. To investigate the influence of growth conditions on the film's properties, a series of samples was prepared with the substrate temperature varied from 225 °C to 675 °C. Following deposition, the films were fabricated into Hall bar devices. Standard photolithography and argon ion milling were used to pattern 8-μm-wide channels, which were aligned along various in-plane crystal directions to probe anisotropic effects. Finally, Ti/Au electrodes were deposited to form the electrical contacts for transport measurements.

*Electrical Transport measurements.* Current-induced magnetization switching was measured by applying DC current pulses with a duration of 30 μs. The resulting magnetization state was read out after a delay by measuring the anomalous Hall resistance with a small sensing AC current. The crystal in-plane Hall effect was characterized by measuring the Hall resistance while sweeping an external magnetic field applied parallel to the film plane. All electrical measurements were performed at room temperature. Longitudinal resistivity and resistivity-temperature ($\rho$-$T$) curves were measured using a Quantum Design physical property measurement system (PPMS).

*STEM sample preparation and characterization.* Cross-sectional samples for scanning transmission electron microscopy (STEM) were prepared using a ZEISS Crossbeam 350 focused ion beam (FIB) system. The samples were initially thinned with a 30 kV $Ga^+$ ion beam, followed by a final polishing step with a low-energy 2 kV beam to minimize surface damage. Microstructural characterization was performed using a ThermoFisher Scientific Spectra 300 STEM, operating at an accelerating voltage of 300 kV. The microscope was equipped with a high-brightness X-CFEG source and an S-CORR probe aberration corrector. High-angle annular dark-field (HAADF) images were acquired using a probe convergence angle of 30 mrad and a collection angle range of 46-200 mrad.

**Data availability**

The data that support the findings of this study are available from the corresponding author upon reasonable request.

**Acknowledgements**

The research was supported by the Ministry of Science and Technology of China (Grant No. 2024YFA1410100), the National Natural Science Foundation of China (Grant No. 12474121, No. 12488101), the Shanxi Province Science and Technology Cooperation and Exchange Special Project (202404041101030), the Science and Technology Commission of Shanghai Municipality (Grant Nos. 2019SHZDZX01), the Innovation Program for Quantum Science and Technology (Grant No. 2021ZD0302500), L.L. acknowledges the Xiaomi Young Scholar program and the Yangyang Development Fund.

**Author contributions**

L.L. and Y.Y.: conceived and designed the experiments; J.C.: performed device fabrication, transport measurements, and data analysis with the assistance of H.Z., H.C., Q.S. and C.P.; J.C., Q.S.: performed STEM experiments; X.L., Y.H., D.G., S.W., Y.L., H.Z., C.L., X.Q., R.Z., and J.J.: contributed to data analysis; A.M. and L.W.: contributed to theoretical and data analysis; L.L., Y.Y. and J.C.: wrote the manuscript and all authors contributed to its final version.

**Competing financial interests' statement**

The authors declare no competing financial interests.

Correspondence and requests for materials should be addressed to L.L. (liul21@sjtu.edu.cn) and Y.Y. (yangym1@shanghaitech.edu.cn).

**References**

1. Kurebayashi H, Garcia JH, Khan S, Sinova J, Roche S. Magnetism, symmetry and spin transport in van der Waals layered systems. *Nature Reviews Physics* **4**, 150–166 (2022).

2. Liu Y, Shao Q. Two-dimensional materials for energy-efficient spin–orbit torque devices.

*ACS nano* **14**, 9389–9407 (2020).

3. Železný J, *et al.* Spin-orbit torques in locally and globally noncentrosymmetric crystals: Antiferromagnets and ferromagnets. *Physical Review B* **95**, 014403 (2017).

4. Manchon A, *et al.* Current-induced spin-orbit torques in ferromagnetic and antiferromagnetic systems. *Reviews of Modern Physics* **91**, 035004 (2019).

5. Liu L, Pai C-F, Li Y, Tseng H, Ralph D, Buhrman R. Spin-torque switching with the giant spin Hall effect of tantalum. *Science* **336**, 555–558 (2012).

6. Miron IM, *et al.* Perpendicular switching of a single ferromagnetic layer induced by in-plane current injection. *Nature* **476**, 189–193 (2011).

7. Liu L, *et al.* Symmetry breaking for current-induced magnetization switching. *Applied Physics Reviews* **10**, (2023).

8. MacNeill D, Stiehl G, Guimaraes M, Buhrman R, Park J, Ralph D. Control of spin–orbit torques through crystal symmetry in WTe2/ferromagnet bilayers. *Nature Physics* **13**, 300–305 (2017).

9. Liu Y, *et al.* Field-free switching of perpendicular magnetization at room temperature using out-of-plane spins from $TaIrTe_4$. *Nature Electronics* **6**, 732–738 (2023).

10. Ma Q, *et al.* Observation of the nonlinear Hall effect under time-reversal-symmetric conditions. *Nature* **565**, 337–342 (2019).

11. Kang K, Li T, Sohn E, Shan J, Mak KF. Nonlinear anomalous Hall effect in few-layer $WTe_2$. *Nature materials* **18**, 324–328 (2019).

12. He P, *et al.* Nonlinear magnetotransport shaped by Fermi surface topology and convexity. *Nat Commun* **10**, 1290 (2019).

13. Xie Q, *et al.* Field-free magnetization switching induced by the unconventional spin–

orbit torque from WTe$_2$. *APL Materials* **9**, (2021).

14. Kao I-H, *et al.* Deterministic switching of a perpendicularly polarized magnet using unconventional spin–orbit torques in WTe$_2$. *Nature materials* **21**, 1029–1034 (2022).

15. Zhang Y, *et al.* Room temperature field-free switching of perpendicular magnetization through spin-orbit torque originating from low-symmetry type II Weyl semimetal. *Science Advances* **9**, eadg9819 (2023).

16. Nye JF. *Physical properties of crystals: their representation by tensors and matrices*. Oxford university press (1985).

17. Johansen Ø, Risinggård V, Sudbø A, Linder J, Brataas A. Current control of magnetism in two-dimensional Fe$_3$GeTe$_2$. *Physical Review Letters* **122**, 217203 (2019).

18. Liu L, *et al.* Symmetry-dependent field-free switching of perpendicular magnetization. *Nat Nanotechnol* **16**, 277–282 (2021).

19. Xue F, Stiles MD, Haney PM. Angular dependence of spin-orbit torque in monolayer Fe$_3$GeTe$_2$. *Physical Review B* **108**, 144422 (2023).

20. García Ovalle D, Pezo A, Manchon A. Spin-orbit torque for field-free switching in C$_{3v}$ crystals. *Physical Review B* **107**, (2023).

21. Vojácek L, *et al.* Field-Free Spin–Orbit Torque Switching in Janus Chromium Dichalcogenides. *Nano letters* **24**, 11889–11894 (2024).

22. Cui Y, *et al.* Antisymmetric planar Hall effect in rutile oxide films induced by the Lorentz force. *Sci Bull (Beijing)* **69**, 2362–2369 (2024).

23. Zhou J, *et al.* Heterodimensional superlattice with in-plane anomalous Hall effect. *Nature* **609**, 46–51 (2022).

24. Liang T, *et al.* Anomalous Hall effect in ZrTe$_5$. *Nature Physics* **14**, 451–455 (2018).

25. Jiang B, *et al.* Chirality-Dependent Hall Effect and Antisymmetric Magnetoresistance in a Magnetic Weyl Semimetal. *Physical Review Letters* **126**, 236601 (2021).

26. Wang Y, *et al.* Antisymmetric linear magnetoresistance and the planar Hall effect. *Nature Communications* **11**, 216 (2020).

27. Peng W, *et al.* Observation of the In-plane Anomalous Hall Effect induced by Octupole in Magnetization Space. Preprint at http://arxiv.org/pdf/2402.15741 (2024).

28. Liu L, *et al.* Crystal Symmetry-Dependent In-Plane Hall Effect. *Nano Lett* **24**, 733–740 (2024).

29. Liu L, *et al.* Current-induced self-switching of perpendicular magnetization in CoPt single layer. *Nat Commun* **13**, 3539 (2022).

30. Tyson TA, Conradson SD, Farrow RF, Jones BA. Observation of internal interfaces in $Pt_xCo_{1-x}$ (x≈0.7) alloy films: A likely cause of perpendicular magnetic anisotropy. *Phys Rev B Condens Matter* **54**, R3702–R3705 (1996).

31. Maranville BB, Schuerman M, Hellman F. Simulation of clustering and anisotropy due to Co step-edge segregation in vapor-deposited $CoPt_3$. *Physical Review B* **73**, (2006).

32. Cross JO, Newville M, Maranville BB, Bordel C, Hellman F, Harris VG. Evidence for nanoscale two-dimensional Co clusters in $CoPt_3$ films with perpendicular magnetic anisotropy. *J Phys Condens Matter* **22**, 146002 (2010).

33. Charilaou M, Bordel C, Berche PE, Maranville BB, Fischer P, Hellman F. Magnetic properties of ultrathin discontinuous Co/Pt multilayers: Comparison with short-range ordered and isotropic $CoPt_3$ films. *Physical Review B* **93**, (2016).

34. Garello K, *et al.* Symmetry and magnitude of spin–orbit torques in ferromagnetic heterostructures. *Nature nanotechnology* **8**, 587–593 (2013).

35. Avci CO, *et al.* Interplay of spin-orbit torque and thermoelectric effects in ferromagnet/normal-metal bilayers. *Physical Review B* **90**, 224427 (2014).

36. Yu G, *et al.* Switching of perpendicular magnetization by spin-orbit torques in the absence of external magnetic fields. *Nat Nanotechnol* **9**, 548–554 (2014).

37. Pai C-F, Mann M, Tan AJ, Beach GSD. Determination of spin torque efficiencies in heterostructures with perpendicular magnetic anisotropy. *Physical Review B* **93**,  (2016).

38. Tao Y, *et al.* Field-free spin–orbit torque switching in L1-FePt single layer with tilted anisotropy. *Applied Physics Letters* **120**,  (2022).

39. Jiang M, Yang X, Qu S, Wang C, Ohya S, Tanaka M. Field-Free Spin-Orbit Torque Magnetization Switching in a Perpendicularly Magnetized Semiconductor (Ga,Mn)As Single Layer. *ACS Appl Mater Interfaces* **16**, 23497–23504 (2024).

40. Han K, Lee KJ, Lee S, Liu X, Dobrowolska M, Furdyna JK. Field-free spin-orbit-torque switching of a single ferromagnetic layer with fourfold in-plane magnetic anisotropy. *APL Materials* **11**,  (2023).

41. Zheng ZY, *et al.* Field-free spin-orbit torque-induced switching of perpendicular magnetization in a ferrimagnetic layer with a vertical composition gradient. *Nature Communications* **12**,  (2021).

42. Jo Y, *et al.* Field-Free Spin-Orbit Torque Magnetization Switching in a Single-Phase Ferromagnetic and Spin Hall Oxide. *Nano Lett* **24**, 7100–7107 (2024).

43. Horiuchi H, *et al.* Single-Layer Spin-Orbit-Torque Magnetization Switching Due to Spin Berry Curvature Generated by Minute Spontaneous Atomic Displacement in a Weyl Oxide. *Adv Mater* **37**, e2416091 (2025).

44. Yang L, *et al.* Field-Free Magnetization Switching in A1 CoPt Single-Layer Nanostructures for Neuromorphic Computing. *ACS Applied Nano Materials* **6**, 9728–

9735 (2023).

45. Li J, *et al.* Interface Effect on the Out‐of‐Plane Spin‐Orbit Torque in the Ferromagnetic CoPt Single Layers. *Advanced Functional Materials* **34**, (2024).